\documentstyle[preprint,aps,version2]{revtex}
\begin{document}
\vspace{-80ex}
\begin{flushright}
NSF-ITP-94-49\\
\vspace{-3mm}
TUIMP-TH-94/59\\
\vspace{-3mm}
NUHEP-TH-94-10\\
\vspace{-2mm}
\it{May, 1994}
\end{flushright}
\vspace{-3mm}

\draft
\begin{title} {\bf On The Spin-Dependent Potential Between \\
Heavy Quark And Antiquark}
\end{title}
\bigskip
\author{Yu-Qi Chen\cite{chen}$^{a,b,d}$
Yu-Ping Kuang\cite{kuang}$^{ a,c,d}$ and
Robert J. Oakes\cite{oakes}$^{ e}$ }

\begin{instit}
$^a$China Center of Advanced Science and Technology (World Laboratory),\\
      P. O. Box 8730, Beijing 100080, China\\
$^b$Institute for Theoretical Physics, University of California, Santa Barbara,
  CA 93106-4030\\
$^{c}$Institute of Modern Physics, Tsinghua University, Beijing 100084,
         China\cite{mailing} \\
$^{d}$Institute of Theoretical Physics, Academia Sinica, Beijing 100080,
China\\
$^e$Department of Physics and Astronomy, Northwestern University, Evanston,
IL 60208
\end{instit}

\bigskip
\centerline{ Abstract}
\vspace{-10mm}
\begin{abstract}
A new formula for the heavy quark-antiquark spin dependent potential
is given by using the techniques developed in the heavy quark effective
theory. The leading logarithmic quark mass terms emerging from the loop
contributions are explicitly extracted and summed up. There is no
renormalization scale ambiguity in this new formula. The spin-dependent
potential in the new formula is expressed in terms of three independent
color-electric and color-magnetic field correlation functions, and it
includes both the Eichten-Feinberg's formula and the one-loop QCD result
as special cases.
\end{abstract}

\narrowtext
\newpage

The study of the structure of the spin-dependent (SD) interaction potential
between heavy quark and antiquark from QCD is one of the interesting problems
in heavy quark physics. So far there are mainly two kinds of approaches  in
the literature. The first kind of approach starts from the static limit
(infinitely heavy quark limit) and makes relativistic corrections via the
$1/m$ expansion, where $m$ stands for the heavy-quark mass. The formula for
the SD potential in this approach was first given by Eichten and
Feinberg\cite{1} in which the potential is  expressed in terms of
certain correlation functions of color-electric and color-magnetic fields
weighted by the  Wilson loop factor.  Later Gromes\cite{2} derived an
important relation between the spin-independent (SI) and the SD potentials
from the Lorentz invariance of the total potential and the correlation
functions
 given in Ref.\cite{1},and it supported the intuitive color-electric flux
tube picture of color-electric confinement suggested by
Buchm\"{u}ller\cite{20}.
The second kind of approach is to calculate the SD potential from perturbative
QCD up to one-loop level and put in  the nonperturbative part of the potential
by hand\cite{3,4}. In this approach, certain terms containing $\ln m$ emerge
from the loop contributions. Furthermore, in the case of unequal masses, new
structure of the order-$1/m^2$ spin-orbit coupling containing $\ln m$ arises
in this approach, which is not included in the first kind of
approach\cite{3,4}. It seems that there exists a discrepancy between these
two kinds of approaches\cite{11}. To understand the essence of this discrepancy
and to have  a deeper understanding of the $\ln m$-dependence in the SD
potential has become  an important theoretical problem in heavy quark physics
for a quite long time. In a recent paper\cite{27}, two of us (Chen and Kuang)
derived some general relations between the SI and SD potentials in the spirit
of the $1/m$ expansion by using the technique of the reparameterization
invariance\cite{14,15,17} developed in the heavy quark effective theory (HQET),
which include the Gromes relation and some new useful relations, and lead to
the conclusion that the general structure of the SD potential is the same as
that obtained in the second approach. However, such a simple symmetry argument
does not concern the problem of the $\ln m$-dependence of the potential.

In this paper, we adopt the conventional formulation in the HQET to construct
an effective QCD Lagrangian including both a heavy quark field and a heavy
antiquark field, with which we study the SD interaction potential.
The conventional HQET has proved to be very powerful in studying the
heavy-light quark systems \cite{18,9,14,15,17,7,10} and most of the useful
techniques developed in it can be applied to the present theory. We emphasize
here that we are working at the quark level to study the interaction potential
between heavy quarks rather than studying the bound state quarkonia. The
reason causing the discrepancy between the two approaches can be easily
understood in the effective Lagrangian formalism. When an effective Lagrangian
is constructed by expanding the full theory in terms of $1/m$, the
ultraviolet-behavior of the theory is changed. Therefore,  when the order of
the $1/m$ expansion and the loop integration are exchanged, which is involved
in passing  from the full theory to the effective theory, differences such as
logarithmic quark mass terms can emerge. In order to reproduce the results of
the full theory, one must match the full theory and the effective one. With
this matching condition these logarithmic quark mass dependent terms can be
explicitly extracted in the coefficients of higher dimensional operators and
can be summed up by using the renormalization group equation (RGE). In  the
original work of Eichten and Feinberg\cite{1} the authors expanded the full
heavy fermion propagator in an ``external'' gluon field $A^{\mu}$ in terms
of $1/m$. It is easy to see that each term in their expression corresponds
to insertions of operators in the effective Lagrangian which contribute to
the SD potential up to order $1/m^2$, but with only tree level coefficients.
Therefore, the logarithmic terms are not accounted for. This is the essential
reason that there is a discrepancy of the quark mass dependence between the
Eichten-Feinberg-Gromes (EFG) formula and the one-loop calculation. With
this insight, we can improve the EFG formula by summing the logarithmic
quark mass terms and establish a consistent picture reconciling these two
kinds of approaches in the framework of the effective Lagrangian. To this
end, we need not only to renormalize the effective Lagrangian up to order
of $1/m^2$, but also to consider the mixing of the nonlocal operators with
local four fermion operators when we use the HQET to calculate the heavy
quark-antiquark Green's functions. Then we can follow the methods developed
in Ref.\cite{26,1} and obtain a new formula for  the spin dependent potential
by using the renormalized effective Lagrangian.  The above results of the
two approaches are just two special approximations of our new formula.
Moreover, our new formula is independent of the renormalization scale
parameter $\mu$, so that it does not suffer from the scale ambiguity as
the second approach does.

Let us first construct  the renormalized effective Lagrangian for one
heavy quark field up to order of $1/m^2$. We start from the full QCD
Lagrangian. As in the conventional HQET, we define the heavy quark field
$h_{v+}(x)$ and the heavy antiquark field $h'_{v-}(x)$  related to the
original field $\psi(x)$ as\cite{18}
\begin{equation}
 h_{v+}(x) \equiv  P_+ e^{imv\cdot x}\psi (x), ~~
 h'_{v-}(x) \equiv P_-e^{-imv\cdot x}\psi (x),
\end{equation}
where $v$ is the velocity of the heavy quark, and
$P_{\pm}\equiv \displaystyle\frac{1\pm \rlap/{v}}{2}$.
Integrating out the quantum fluctuation of the quark field,
we obtain the heavy quark effective Lagrangian ${\cal L}_c$\cite{9,17}.
After expanding order by order in powers of $1/m$, the first three terms are
\begin{eqnarray}
{\cal L}_0 &=&c_0 \bar{h}_{v+}(x) iD\cdot v h_{v+}(x),\label{l0}\\
{\cal L}_1 &=&c_0c_1 \bar{h}_{v+}(x) \displaystyle\frac{(iD)^2}{2m} h_{v+}(x)-
c_0c_2\bar{h}_{v+}(x) \displaystyle\frac{(iD\cdot v)^2}{2m} h_{v+}(x) +
c_0c_3g_s\bar{h}_{v+}(x) \displaystyle\frac{G_{\mu\nu} \sigma^{\mu\nu} }{4m}
h_{v+}(x), \label{l1}\\
{\cal L}_2 &=&
c_0\displaystyle\frac{g_s}{4m^2} \bar{h}_{v+}(x) (c_4v^\nu D^\mu G_{\mu\nu} +
ic_5 \sigma^{\mu\nu} v^\sigma D_\mu G_{\nu\sigma} ) h_{v+}(x)\nonumber\\
 && -\displaystyle\frac{c_0}{4m^2}
\bar{h}_{v+}(x)\left[ c_6(iD)^2 -ic_7(D\cdot v)^2
-c_8\displaystyle\frac{g_s}{2} G_{\mu\nu} \sigma^{\mu\nu} \right]  iD\cdot v
 h_{v+}(x),\label{l2}
\end{eqnarray}
where $iD_{\mu} =i\partial_\mu-gA_{\mu}^a T^a$. The last term in
${\cal L}_2$ has no contribution in order $1/m^2$ due to the equation of
motion. At tree level the $c_i$'s are all unity. Note that after the
expansion the high energy behavior is different from that in the full theory.
So the operators in (\ref{l0})-(\ref{l2}) need to be renormalized and their
coefficients can be determined by matching to the full theory. Here
$\sqrt{c_0}$ corresponds to the wavefunction renormalization constant, and
$c_1-c_3$ have been calculated in Refs.\cite{7} and \cite{10}, using the RG
summation, Ref.\cite{10} gives
\begin{equation}
c_1(\mu,m)=1, ~~
c_2(\mu,m)=3\left(\displaystyle\frac{\alpha_s(\mu)}
{\alpha_s(m)}\right)^{-{\frac{8}{25}}} -2, ~~
c_3(\mu,m)= \left(\displaystyle\frac{\alpha_s(\mu)}
{\alpha_s(m)}\right)^{-{\frac{9}{25}}},
\label{c13}
\end{equation}
where $c_i\equiv c_i(\mu,m)$. The coefficients $c_4-c_6$  can be determined
by the reparameterization invariance of ${\cal L}_c$ which
leads to the reparameterization invariance of the renormalized Lagrangian.
This is shown as follows: Consider the infinitesimal velocity transformation
$v\to v+\Delta v$\cite{17}. The infinitesimal transformation of
$\delta {\cal L}$ can be written as
\begin{equation}
\delta {\cal L}= \delta{\cal T}_0 + {1\over 2m} \delta{\cal T}_1,
\end{equation}
where
\begin{equation}
\delta{\cal T}_0= c_0(1-c_1) \bar{h}_{v+}(x) iD\cdot  \Delta\!v h_{v+}(x),
\label{t0}
\end{equation}
and
\begin{equation}
\begin{array}{lcl}
\delta{\cal T}_1 &=& c_0(1-2 c_2+c_6)\;
\bar{h}_{v+}(x) iD\cdot \Delta v iD\cdot v  h_{v+}(x)\\
 &+& c_0(c_3-c_2+{1\over 2} c_4- {1\over 2} c_5)\;
 \bar{h}_{v+}(x) \gamma^\mu v^\nu igG_{\mu\nu} {\;\rlap/{\!\!{\Delta}\!
v}\over 2} h_{v+}(x)\\
 &+&  c_0(1-c_2-c_3+{1\over 2} c_4+ {1\over 2} c_5)\;
\bar {h}_{v+}(x) {\;\rlap/{\!\!{\Delta}\! v}\over 2}
\gamma^\mu v^\nu igG_{\mu\nu} h_{v+}(x).
\label{t1}
\end{array}
\end{equation}
Different terms in (\ref{t0})-(\ref{t1}) are of different Lorentz structures,
and therefore, they should vanish separately. $\delta{\cal T}_0=0$ leads to
$c_1=1$, i.e., the kinetic energy term is not renormalized\cite{15}, and
$\delta{\cal T}_1=0$ gives the two relations
\begin{equation}
 c_4 = c_6= 2c_2-1, ~~~~~~
 c_5 = 2c_3-1.
\label{c45}
\end{equation}
Using Eq.(\ref{c13}) we obtain
\begin{equation}
c_4(\mu,m)=6\left(\displaystyle\frac{\alpha_s(\mu)}
{\alpha_s(m)}\right)^{-{8\over 25}} -5, ~~~~~~
c_5(\mu,m)= 2\left(\displaystyle\frac{\alpha_s(\mu)}
{\alpha_s(m)}\right)^{-{9\over 25}}-1.
\end{equation}

The effective Lagrangian for antiquark field can be obtained by simply
replacing $v$ by $-v$ and $h_{+v}(x)$ by $h'_{-v}(x)$  in the above effective
Lagrangian.

To study the heavy quark-antiquark interaction potential we are going
to evaluate the heavy quark-antiquark four point Green's function to order
$1/m^2$. In doing this, we need to calculate all the possible order $1/m$
operator insertions. Then, additional divergences will appear from double
insertions of these operators, i.e., these bilocal operators will mix with
certain local four-fermion operators. Let us consider a general unequal mass
case, e.g. the heavy quark $Q_1$ with mass $m_1$ and the antiquark $\bar{Q_2}$
with mass $m_2$, as in the case of the $c\bar{b}$. To order $1/m^2$, there are
only two local dimension 6 color singlet four fermion operators $O_1(x)$ and
$O_2(x)$:
\begin{eqnarray}
O_1(x)&=& \displaystyle\frac{g_s^2}{4m_1m_2} \bar{h}_{v+i}(x)
\sigma^{\mu\nu} h_{v+j}(x) \bar{h'}_{v-j}(x) \sigma_{\mu\nu} h'_{v-i}(x), \\
O_2(x)&=& \displaystyle\frac{g_s^2}{4m_1m_2} \bar{h}_{v+i}(x)
\sigma^{\mu\nu} h_{v+i}(x) \bar{h'}_{v-j}(x) \sigma_{\mu\nu} h'_{v-j}(x),
\end{eqnarray}
 where $i,j=1,2,\cdots,N_c$ are color indices. Suppose $m_1 >m_2$. The
heavy quark antiquark effective Lagrangian up to order $1/m_1^2$,
$1/m_1m_2$, and $1/m_2^2$ can be constructed in two steps as follows.
Starting from the Lagrangian in the full theory, we first treat $Q_1$
as a heavy quark, and obtain
\begin{equation}
{\cal L'}={\cal L}_{Q_1eff} + {\cal L}_{Q_2}.
\label{hl}
\end{equation}
We do not need to add new operators in (\ref{hl}) because there are no
divergent terms of the form of $O_1(x)$ and $O_2(x)$. Next, we treat $Q_2$
as a heavy antiquark, and obtain
\begin{equation}
{\cal L^{''}}={\cal L}_{Q_1eff} + {\cal L}_{Q_2eff} +d_1(\mu) O_1(\mu)
+d_2(\mu) O_2(\mu),
\label{leff}
\end{equation}
where the last two terms are the two necessary dimension-6 operators
with unknown coefficients $d_1(\mu)$ and $d_2(\mu)$, respectively.
Now we  determine $d_1(\mu)$ and $d_2(\mu)$  by using the RGE.  It is easy
to see that only the magnetic operator insertion in each fermion line will
mix with $O_1$ and $O_2$ due to the Lorentz structure. Let us denote the
resulting contribution by
\begin{equation}
O_0(x)\equiv \displaystyle\frac{g_s^2}{16m_1m_2} \int d^4y T^*\left[
\bar{h}_{v+}(x) {G_{\mu\nu} \sigma^{\mu\nu} } h_{v+}(x)
\bar{h}'_{v-}(y) {G_{\alpha\beta} \sigma^{\alpha\beta} } h'_{v-}(y) \right],
\end{equation}
and its coefficient by $d_0(\mu)$. Here  $T^*$ means time ordering. The
coefficients $d(\mu)\equiv (d_0(\mu),~ d_1(\mu),~ d_2(\mu))$ satisfy the
renormalization group equation
\begin{equation}
\mu\displaystyle\frac{d}{d\mu}d(\mu) +d(\mu)\gamma=0,
\label{rg}
\end{equation}
where $\gamma$ is the anomalous dimension matrix. A straightforward
one-loop calculation using the HQET technique gives
\begin{equation}
\gamma=\displaystyle\frac{g^2}{4\pi^2}
\left(
\begin{array}{ccc}
-N_c & \displaystyle\frac{N_c}{8} & -\displaystyle\frac{1}{8} \\
 0 & 0 & 0 \\
 0  & 0 & 0
\end{array}
\right).
\end{equation}
Here $\gamma_{00}=2\gamma_{mag}$. The results $\gamma_{10}=\gamma_{20}=0$
means that local operators are not able to mix with  bilocal operators.
However, bilocal operators can mix with local operators, and there are
two box, two cross, and a ``fish'' diagram contributing to $\gamma_{01}$
and $\gamma_{02}$. That $\gamma_{11}=\gamma_{12}=\gamma_{21}=\gamma_{22}=0$
can be understood as follows: The anomalous dimensions are gauge independent.
If we take an axial gauge where $v\cdot A=0$ gluonic interactions decouple from
the fermion field in the zeroth order effective Lagrangian (\ref{l0}). The
initial condition for the RGE is determined by matching the effective theory
to the full theory at $\mu=m_2$, i.e.
\begin{equation}
d(m_2)=\left(~c_3(m_2,m_1),0,0~\right).
\end{equation}
With this, the solution of the RGE (\ref{rg})  are
\begin{equation}
\begin{array}{lcl}
d_0(\mu) &=& c_3(\mu,m_2)c_3(\mu,m_1)=
\left(\displaystyle\frac{\alpha_s^2(\mu)}{\alpha_s(m_1)\alpha_s(m_2)}
\right)^{-{9\over 25}},\\
d_1(\mu) &=& \displaystyle\frac{1}{8}  c_3(m_2,m_1)[1-c_3^2(\mu,m_2)]
=\displaystyle\frac{1}{8}
\left(\displaystyle\frac{\alpha_s(m_2)}{\alpha_s(m_1)}\right)^{-{9\over 25}}
\left[ 1-\left(\displaystyle\frac{\alpha_s(\mu)}{\alpha_s(m_2)}\right)^
{-{18\over 25}} \right], \\
d_2(\mu) &=&-\displaystyle\frac{1}{N_c}d_1(\mu).
\end{array}
\end{equation}
Our effective Lagrangian (\ref{leff}) is thus completely determined.

Now we apply it to calculate spin-dependent force.  We  shall take
$v=(1,0,0,0)$ and denote $h_{v+}(x)$, $h'_{v-}(x)$ as $h(x)$, $h'(x)$
for short. Similar to Ref.\cite{1}, we introduce a gauge invariant four-point
Green's function
\begin{equation}
I=\langle  0| T^*[\bar{h}'(y_2)\bar{\Gamma}_B P(y_2,y_1) h(y_1)]\,
[\bar{h}(x_1){\Gamma}_A P(x_1,x_2) h'(x_2)]\,|0 \rangle,
\end{equation}
where $P(x,y)\equiv P\exp \left[ ig\int^x_y dz_\mu A^\mu(z)\right]$
is the path-ordered exponential\cite{24,25}.
As is argued in Refs.\cite{26,1}, in the limit that the time interval
$T\equiv (y_1^0+y_2^0)/2-(x_1^0+x_2^0)/2 \to \infty $,
with $x_2^0-x_1^0$ and $y_2^0-y_1^0$ fixed, the limit of $I$ is
\begin{equation}
I\to \delta_{AB} \delta(\vec{r}_x-\vec{r}_y) \exp [-T\epsilon(r)],
\end{equation}
where $\vec{r}_x=\vec{x}_1-\vec{x}_2$, $\vec{r}_y=\vec{y}_1-\vec{y}_2$,
$r=|\vec{r}_x|$, and $\epsilon(r)$ is just the static energy between the quark
and antiquark separated by the  spatial distance $r$. Here  the appropriate
ordering of the limits is that first $m\to \infty $ and then $T\to \infty$,
so that the motion of the quark and the antiquark can be treated
perturbatively\cite{26,1}.

Taking all the operators with dimension higher than 4 in  the Lagrangian
as the perturbative part, $I$ can be calculated by using  standard
perturbation theory. In the calculation, the zeroth order full fermion
propagator $S_0(x,y,A)$ in the external gluon field $A^{\mu}$ is used and
it is\cite{26,1}
\begin{equation}
S_0(x,y,A)=-i\theta\ (x_0-y_0)P(x_0,y_0)\delta(\vec{x}-\vec{y}).
\end{equation}
Next we define   the symbol $\langle\cdots\rangle$ and $\tilde{I}$ as
\begin{equation}
\langle Q(x)\rangle \equiv \int [dA^\mu]Tr \left\{ P\left[\exp
\left(ig\oint_{C(r,T)} dz_\mu A^\mu(z) \right)
Q(x)\right]\right\}_{x\in C} \exp (iS_{YM}(A)),
\end{equation}
\begin{equation}
I\equiv Tr (P_+ \bar{\Gamma} P_-\Gamma \tilde{I} )\delta(\vec{x}_1-\vec{y}_1)
\delta(\vec{x}_2-\vec{y}_2).
\end{equation}
To order $1/m_1^2$, $1/(m_1m_2)$ and $1/m_1^2$, $\tilde{I}$ can be
expressed as
\begin{equation}
\begin{array}{lcl}
\tilde{I}&=&\langle 1\rangle  +i\left[ \displaystyle\int^{T/2}_{T/2}dz
\displaystyle\frac{1}{m_1}\langle {\rm\bf D}^2({\rm\bf x}_1,z)-
c_3(\mu,m_1)g_s(\mu)\vec{\sigma}_1\cdot B({\rm\bf x}_1,z)\rangle  +
(1\leftrightarrow 2) \right] \\[5mm]
&-&\displaystyle\frac{g_s(\mu)}{4m_1^2}\displaystyle\int^{T/2}_{T/2}
dz[(c_4(\mu,m_1)\delta_{ij}
-c_5(\mu,m_1)i\epsilon_{ijk}\sigma^k_1)\langle E^i({\rm\bf x}_1,z)
D^j({\rm\bf x}_1,z)\rangle + (1\leftrightarrow 2) \\[5mm]
&-& \displaystyle\frac{1}{m_1^2}\displaystyle\int^{T/2}_{-T/2}dz
\displaystyle\int^{T/2}_{-T/2}dz'\theta(z'-z) [\langle ({\rm\bf D}^2
-c_3(\mu,m_1)g_s(\mu)\sigma_1^iB^i)({\rm\bf x}_1,z)\\[5mm]&& ({\rm\bf D}^2
-c_3(\mu,m_1)g_s(\mu)\sigma_1^iB^i)
({\rm\bf x}_1,z')\rangle  +(1\leftrightarrow 2)]
-\displaystyle\frac{1}{m_1m_2}\displaystyle\int^{T/2}_{T/2}dz
\displaystyle\int^{T/2}_{-T/2}dz'\\[5mm]
&& \langle ({\rm\bf D}^2-c_3(\mu,m_1)g_s(\mu)\sigma_1^iB^i)({\rm\bf x}_1,z)
({\rm\bf D}^2-g_s(\mu)c_3(\mu,m_2)\sigma_2^jB^j)({\rm\bf x}_2,z')\rangle
\\[5mm]
&+&\displaystyle\frac{N_cg_s^2(\mu)}{2m_1m_2}T d(\mu)\sigma_1^i\sigma_2^i
\delta^3({\rm\bf x}_1-{\rm\bf x}_2),
\end{array}
\end{equation}
where
\begin{equation}
d(\mu)=d_1(\mu)+\displaystyle\frac{d_2(\mu)}{N_c}.
\end{equation}
Similar to the derivation given in Ref.\cite{1}, we obtain the
spin-dependent potential
\begin{equation}
\begin{array}{lcl}
V(r) &=& V_0(r)
+\left( \displaystyle\frac{{\rm\bf{S}_1 }}{m_1^2}
+\displaystyle\frac{{\rm\bf{S}_2 }}{m_2^2} \right) {\bf \cdot L}
\left[\left(c_+(\mu,m_1,m_2)-\displaystyle\frac{1}{2}\right)\,
\displaystyle\frac{dV_0(r)}{dr} \right.\\[5mm]&+&
 \left. c_+(\mu,m_1,m_2)\displaystyle\frac{dV_1(\mu,r)}{dr}\,\right]
+ \left( \displaystyle\frac{{\rm\bf {S}_1 + {S}_2 }}{m_1m_2} \right)
{\bf \cdot L} c_+(\mu,m_1,m_2)\displaystyle\frac{1}{r}
\displaystyle\frac{dV_2(\mu,r)}{dr} \\[5mm]
&+&\displaystyle\frac{1}{m_1m_2}\displaystyle\frac{({\rm\bf {S}_1}\cdot
{\rm\bf {r}})( {\rm\bf {S}_2}\cdot {\rm\bf {r}})-
\displaystyle\frac{1}{3}{\rm\bf {S}_1}
\cdot{\rm\bf {S}_2}\,r^2 }{r^2}\,
c_3(\mu,m_1)c_3(\mu,m_2)V_{3}(\mu,r) \\[5mm] &+&
\displaystyle\frac{1}{3} \, \displaystyle\frac{1}{m_1m_2}
{\rm\bf{S}_1\cdot {S}_2}
\left[(c_3(\mu,m_1)c_3(\mu,m_2)V_4(\mu,r)- 6N_c g_s^2(\mu)
d(\mu)\delta({\rm\bf r}) \right]
\\[5mm]
&+& \left( \displaystyle\frac{{\rm\bf{S}_1 }}{m_1^2}
-\displaystyle\frac{{\rm\bf{S}_2 }}{m_2^2} \right) {\bf \cdot L}
c_-(\mu,m_1,m_2)\displaystyle\frac{1}{r}\displaystyle\frac{d[V_0(\mu,r)
+V_1(\mu,r)]}{dr} \\[5mm]&+&
 \left( \displaystyle\frac{{\rm\bf {S}_1 - {S}_2 }}{m_1m_2} \right)
{\bf \cdot L} c_-(\mu,m_1,m_2)
\displaystyle\frac{1}{r} \displaystyle\frac{dV_2(\mu,r)}{dr},
\end{array}
\label{vpot}
\end{equation}
where
\begin{eqnarray}
&&c_+(\mu,m_1,m_2)=\displaystyle\frac{1}{2}[c_3(\mu,m_1)+c_3(\mu,m_2)],~~~~
c_-(\mu,m_1,m_2)=\displaystyle\frac{1}{2}[c_3(\mu,m_1)-c_3(\mu,m_2)], \\
&&{\rm and}\nonumber \\
&&V_0(r) \equiv-\lim_{T\to \infty} \displaystyle\frac{\ln\langle 1\rangle }
{T},\\[3mm]
&&r_k\displaystyle\frac{1}{r}\displaystyle\frac{dV_1(\mu,r)}{dr} \equiv
\lim_{T\to\infty} \epsilon_{ijk} \displaystyle\int^{T/2}_{-T/2}dz
\displaystyle\int^{T/2}_{-T/2}dz' \left(\displaystyle\frac{z'-z}{T}\right)
g_s^2(\mu)/2\langle B^i({\rm\bf x}_1,z)E^j({\rm\bf x}_1,z')\rangle /\langle
1\rangle , \\[3mm]
 & &  r_k\displaystyle\frac{1}{r}\displaystyle\frac{dV_2(\mu,r)}{dr} \equiv
\lim_{T\to\infty}
\epsilon_{ijk} \displaystyle\int^{T/2}_{-T/2}dz
\displaystyle\int^{T/2}_{-T/2}dz' \left(\displaystyle\frac{z'}{T}\right)
g_s^2(\mu)/2\langle B^i({\rm\bf x}_2,z)E^j({\rm\bf x}_1,z')\rangle /\langle
1\rangle , \\[3mm]
& & [(\hat{r}_i\hat{r}_j-\displaystyle\frac{\delta^{ij} }{3})V_3(\mu,r)
+\displaystyle\frac{\delta^{ij}}{3}V_4(\mu,r)]
\equiv \lim_{T\to \infty}\displaystyle\int^{T/2}_{-T/2}dz
\displaystyle\int^{T/2}_{-T/2}dz'
\displaystyle\frac{g_s^2(\mu) }{T}
\langle B^i({\rm\bf x}_1,z)B^j({\rm\bf x}_2,z')\rangle /\langle
1\rangle .
\end{eqnarray}

In Ref.\cite{2}, Gromes derived a relation
$\displaystyle\frac{d}{dr}[V_0(r)+V_1(\mu,r)-V_2(\mu,r)]=0.$
The relation $c_5(\mu,m)=2c_3(\mu,m)-1$ in (\ref{c45}) obtained from
reparameterization invariance ensures that the general relations derived
in Ref. \cite{27} are all satisfied. It also shows that those relations are
consistent with each other. Using those relation, the SD potential can be
simplified as
\begin{equation}
\begin{array}{lcl}
V(r) &=& V_0(r)
+\left( \displaystyle\frac{{\rm\bf{S}_1 }}{m_1^2}
+\displaystyle\frac{{\rm\bf{S}_2 }}{m_2^2} \right) {\bf \cdot L}
\left(c_+(\mu,m_1,m_2)\displaystyle\frac{dV_2(\mu,r)}{dr}
-\displaystyle\frac{1}{2}\displaystyle\frac{dV_0(r)}{dr}\right)  \\[5mm] &+&
\left( \displaystyle\frac{{\rm\bf {S}_1 + {S}_2 }}{m_1m_2} \right)
{\bf \cdot L} c_+(\mu,m_1,m_2)\displaystyle\frac{1}{r}\displaystyle
\frac{dV_2(\mu,r)}{dr}  \\[5mm]
&+&\displaystyle\frac{1}{m_1m_2}\displaystyle\frac{({\rm\bf {S}_1}\cdot
{\rm\bf {r}})( {\rm\bf {S}_2}\cdot
{\rm\bf {r}})-\displaystyle\frac{1}{3}{\rm\bf {S}_1}
\cdot{\rm\bf {S}_2}\,r^2 }{r^2}\,
c_3(\mu,m_1)c_3(\mu,m_2)V_{3}(\mu,r) \\[5mm] &+&
\displaystyle\frac{1}{3} \, \displaystyle\frac{1}{m_1m_2} {\rm\bf{S}_1\cdot
{S}_2} \left[(c_3(\mu,m_1)c_3(\mu,m_2)V_4(\mu,r)
-6N_c g_s^2(\mu) d(\mu)\delta({\rm\bf r})\right]
\\[5mm] &+&
\left[\left( \displaystyle\frac{{\rm\bf{S}_1 }}{m_1^2}
-\displaystyle\frac{{\rm\bf{S}_2 }}{m_2^2} \right) {\bf \cdot L}
+ \left( \displaystyle\frac{{\rm\bf {S}_1 - {S}_2 }}{m_1m_2} \right)
{\bf \cdot L}\right] c_-(\mu,m_1,m_2)
\displaystyle\frac{1}{r} \displaystyle\frac{dV_2(\mu,r)}{dr}.
\end{array}
\label{spot}
\end{equation}
This is our new formula for the spin-dependent quark-antiquark potential.

In Eq.~(\ref{spot}), if we take each coefficient to be its tree level value,
i.e., $c_3(\mu,m)=c_+(\mu,m_1,m_2)=1$ and  $c_-(\mu,m_1,m_2)=d(\mu)=0$,
our result reduces to the EFG formula. Next we compare our result with that
in the one-loop QCD calculation. First we see that $V_5(\mu,m_1,m_2)$
introduced in Ref. \cite{4} is not an independent function. If we take
the one-loop values of $c_3(\mu,m)$, $c_\pm(\mu,m_1,m_2)$, $d(\mu)$, and then
calculate the correlation functions to one-loop, our formula (\ref{spot})
reproduces all the logarithmic mass terms in Refs.\cite{3,4}.

In conclusion,  the leading logarithmic quark mass terms emerging from
the loop contributions are explicitly extracted and summed up by matching
the effective theory and the full theory and solving the renormalization
group equation. The discrepancy appearing in the EFG results and one-loop
calculation can then be understood. Our result  shows that the effective
theory can reproduce the full theory beyond tree level in $1/m^2$ and can
be used in calculating the Green's functions with two heavy quark external
lines.

One of the authors, Y.-Q. Chen, would like to thank Mark Wise for
instructive discussions on this work. He also would like to thank
David Kaplan, Lisa Randall and Michael Luke for useful discussions.
This work is supported partly by the  National Natural Science
Foundation of China, the Fundamental Research Foundation of Tsinghua
University, and the U.S. Department of Energy, Division of High Energy
Physics, under Grant DE-FG02-91-ER40684.

\end{document}